\documentstyle[prl,aps,twocolumn,epsf]{revtex}

\begin{document}
\title{A Proof of Duality in Structure Functions Near $x = 1$}  

\author{
Geoffrey B. West\footnote{gbw@pion.lanl.gov}
}

\address{
Theoretical Division, T-8, MS B285,
Los Alamos National Laboratory,
Los Alamos, NM  87545}


\maketitle                 
\begin{abstract}
{A proof of Bloom-Gilman duality which relates an integral over the low-mass
resonances in deep inelastic structure functions to an integral over the
scaling region near $x = 1$ is given. It is based on general analytic
properties of the corresponding virtual Compton amplitude but is insensitive to
its asymptotic behaviour.}
\end{abstract}

\vskip1cm

\narrowtext

Renewed interest in the large $x(\approx 1)$ behaviour of
deep inelastic structure functions has been driven in large part by new data from
SLAC \cite{note}\cite{SLAC}. In addition, a wealth of detailed data can be expected
from CEBAF in the near future, albeit in the scaling transition region at
relatively modest values of $Q^2$. Together these will refocus attention on
Bloom-Gilman duality, which connects an integral over the resonant contributions
below the onset of scaling to an integral over the
threshold scaling region above the resonances \cite{BG}. This is
closely connected to the ``inclusive-exclusive" connection which
 relates the $x\approx 1$ behaviour of
the scaling structure functions to the large $Q^2$ behaviour of the resonance
form factors and which was first derived from the quark-parton model \cite{gbw}\cite{brod}. As
originally formulated, duality can be expressed in the following way:  
\begin{equation}  
{2M\over Q^2}\int_0^{\bar\nu}d\nu F_2(\nu ,Q^2)
            = \int_1^{\bar\omega '} d\omega ' F_2(\omega ')
\label{one}
\end {equation}
Here, $Q^2$ and $\nu$ are to be taken to be sufficiently large that
$F_2(\nu,Q^2)$, the conventional structure function, scales, i.e., it becomes a
function only of  $\omega ' \equiv (2M\nu + M_0^2)/Q^2 = 1/x + M_0^2/Q^2$ with
$M_0$ an ``arbitrary" scale of order $M$. In the original work $F_2$ was
assumed to scale exactly; below we shall discuss the inclusion of logarithmic
scaling violations dictated by asymptotic freedom. The limit on the integral in
(\ref{one}), $\bar\nu$, defines the transition from the non-scaling resonance
region to the continuum scaling region; furthermore, $\bar\omega ' = (2M\bar\nu +
M_0^2)/Q^2$. Bloom and Gilman found that the data satisfies this sum rule rather
well if $M_0$ is identified with $M$. Thus, in some average sense as specified by
Eq.~(\ref{one}), the scaling function smoothly interpolates through the bumpy
resonant region. They proposed a derivation of this finite energy sum rule using a
superconvergence relation which followed from some assumptions about the high
energy (i.e., large $\nu$) behaviour of $F_2$. The form of this asymptotic
behaviour was guided by conventional Regge phenomenology for the corresponding
$q^2 = 0$ Compton amplitude and included the presumed absence of fixed
poles. If present, these would contribute only to its real part and, in their derivation,
would lead to an unknown contribution to
Eq.~(\ref{one}). In addition, the asymptotic behaviour was assumed to hold uniformly from
$q^2 = 0$ into the scaling region  where $q^2$ becomes large and space-like. As such, it is, in
principle, sensitive to the small-$x$ behaviour of $F_2$ which is still a matter of
conjecture, both theoretically and experimentally. 

With this in mind, we revisit the problem and give a relatively 
simple proof of
duality which is essentially independent of both the high energy and fixed pole
behaviour of the Compton amplitude. The former enters only in as much as it
determines the number of subtractions required for a fixed-$q^2$ dispersion
relation in $\nu$ to converge. These play no role in the proof which is trivially amended if
the number of subtractions is different from what has been perenially assumed. The full Compton
amplitude, $T_2(q^2,\nu)$, whose imaginary part is proportional to $F_2(q^2,\nu)$ is, for
fixed $q^2(\le 0)$, an analytic function of $\nu$ except for cuts along the real
axis; (see Fig.~1).

\bigskip
\begin{figure}
\narrowtext
\epsfxsize=\hsize\epsfbox{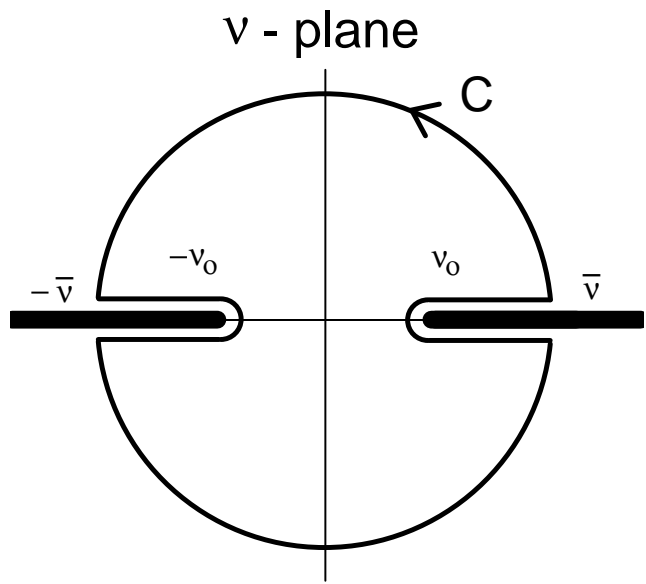}
\medskip
\caption{The complex $\nu$-plane showing the contour $C$ wrapping around the cuts beginning at the threshold, $\pm\nu_0$.}
\end{figure}
%


Using crossing symmetry this can be expressed in the form of the
well-known dispersion relation:
\begin{equation}
T_2(q^2,\nu) = \int^{\infty}_0 {d{\nu '}\over {{\nu '}^2 -
\nu^2}}F_2(q^2,\nu ') 
\label{two}
\end{equation}
With a change of variables this can be expressed as follows:
\begin{eqnarray}
\nu T_2(q^2,\omega ')&\nonumber\\ \span\displaystyle\hspace{0.02cm} = (\omega ' -{M_0^2\over q^2})
                           \int^{\infty}_{1+{M_0^2\over q^2}} d\omega ''
                          { F_2(q^2,\omega '')\over
                          {(\omega '' - \omega ')
                           (\omega '' + \omega ' - 2{M_0^2\over q^2})}}
\label{three}
\end{eqnarray}
When $Q^2\gg M_0^2$ this reduces to
\begin{equation}
\nu T_2(q^2,\omega ') = \omega '
                           \int^{\infty}_1 d\omega ''
                          { F_2(q^2,\omega '')\over
                          {{\omega ''}^2 - {\omega '}^2}}
\label{four}
\end{equation}
Following Bloom and Gilman we will assume that scaling (i.e., $F_2(q^2,\omega ')\approx
F_2(\omega ')$) sets in when $Q^2\approx \bar Q^2 \ge M_0^2$ and $s\approx \bar s$
where $\bar s$ exceeds the mass-squared of the last resonance. Note that 
$\bar\nu = (\bar s - M^2 + Q^2)/2M$ with $Q^2 > \bar Q^2$.

Consider now the contour, C, shown in Fig. 1 consisting of a
circle of radius $\bar\nu$ together with line integrals around the cuts from threhold 
$\nu = \pm\nu_0$  ($\equiv Q^2/2M$) to $\pm\bar\nu$. Then, since it encloses no
singularities, 
\begin{eqnarray}
\int^{\bar\nu}_{\nu_0} F_2(q^2,\nu)d\nu = \oint_{\bar\nu} \nu T_2(q^2,\nu){d\nu\over 2\pi i}\nonumber\\
                                        = \bar\nu\int_0^{2\pi}{d\theta\over 2\pi i} \bar\nu
                                           T_2(q^2,\bar\nu
                                           e^{i\theta}) e^{i\theta}
\label{five}
\end{eqnarray}
where the contour is the circle of radius 
$\vert\bar\nu\vert$ centred at the origin. Now, if $Q^2\ge\bar Q^2$, then
this relates an integral over the resonances {\it below} the scaling threshold to the
full Compton amplitude evaluated at $\bar\nu$ {\it in the scaling region above the
resonances}. The left-hand-side is just that which occurs in
the duality relationship, Eq.~(\ref{one}). To derive (\ref{one})  we can use the 
representation Eq.~(\ref{four}) in the right-hand-side of  Eq.~(\ref{five}) to write 
\begin{equation}
\int^{\bar\nu}_{\nu_0} F_2(q^2,\nu)d\nu ={q^2\over 2M}\oint_{\bar\omega '} d\omega '\omega '
                           \int^{\infty}_1 {d\omega ''\over 2\pi i}
                          { F_2(q^2,\omega '')\over
                          {{\omega ''}^2 - {\omega '}^2}}
\label{six}
\end{equation}
The contour integral on the right-hand-side is to be evaluated around a circle of
radius $\bar\omega '$ centred at the origin of the complex $\omega '$-plane as shown
in Fig. 2. 


Since the integrals are absolutely convergent the order of integration can be
interchanged. Furthermore, it is easy to see that 
\begin{equation}
{1\over 2\pi i}\oint_{\bar\omega '} {d\omega '\omega '\over{{\omega ''}^2 - {\omega '}^2}}
                                             = \theta (\bar\omega ' - \omega '')
\label{seven}
\end{equation}
Using this in Eq.~(\ref{six}) immediately leads to Eq.~(\ref{one})
\cite{finite}:
\begin{equation}  
{2m\over q^2}\int_0^{\bar\nu}d\nu F_2(\nu ,q^2)
            = \int_1^{\bar\omega '} d\omega ' F_2(q^2,\omega ')
\label{eight}
\end {equation}
Notice that no assumption about either the high energy (i.e., large $\nu$) or the 
large
$\omega '$ (i.e., small $x$) behaviours of $T_2$ need be made to derive this nor 
do we need to assume exact
scaling. Furthermore, the presence of fixed poles, or unknown real parts,
contributing to $T_2$ play no role since they are analytic. 

\begin{figure}
\narrowtext
\epsfxsize=\hsize\epsfbox{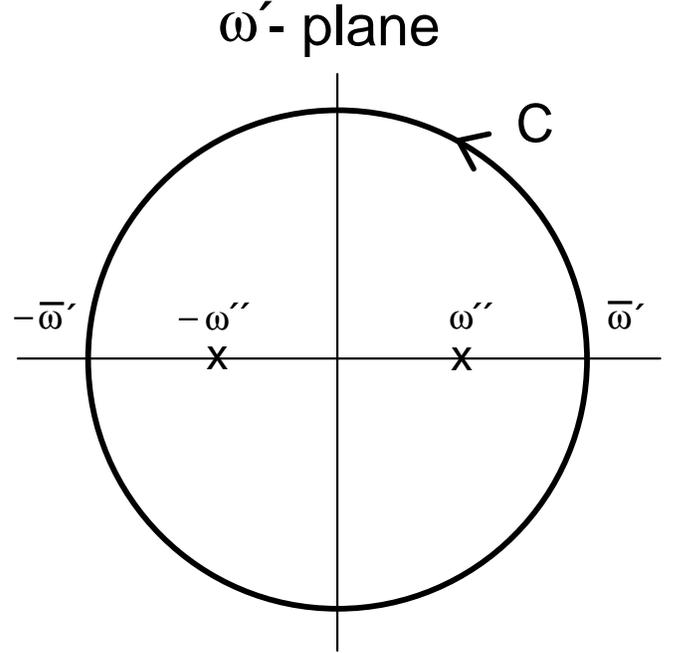}
\medskip
\caption{The complex $\omega '$-plane showing the circular contour of radius $\bar\omega '$ used in 
and the poles at $\pm\omega ''$.}
\end{figure}

Some further remarks
are in order:

i) Similar sum rules can be straightforwardly derived for the transverse structure function
$F_1(q^2,\omega ')$.  Although the dispersion relation for the corresponding Compton amplitude,
$T_1(q^2,\nu)$, requires (at least) one subtraction this will not contribute to the sum rule
since it is a purely real contribution. 

ii) In addition, higher moment sum rules can also be derived following the same
procedure; for example, for sufficiently large $Q^2$, we find
\begin{equation}  
({2M\over -q^2})^{2m+1}\int_0^{\bar\nu}d\nu \nu^{2m} F_2(\nu ,q^2)
            = \int_1^{\bar\omega '} d\omega ' F_2(q^2,\omega '){\omega '}^{2m}
\label{ten}
\end {equation}
This is valid provided $Q^2/M_0^2\gg m\ge 0$. This condition can be relaxed by
keeping explicit terms of $O(M_0^2/Q^2)$. For $m<0$ the situation is a little more
subtle since there are singularities (poles at the origin) inside $C$.
Effectively the resulting sum rules turn out to be identities and contain no new
information.

iii) Recall that, if the transition form factor to a given resonance is generically denoted by
$G_r(q^2)$, and if, in the scaling region, $F_2(q^2, \omega ')\approx A(\omega ' - 1)^p$ when
$\omega '\approx 1$ (assuming, for the moment, exact scaling), then saturating
Eq.~(\ref{eight}) with just the resonances leads to the relationship (valid only for large
$Q^2$) 
\begin{equation}
\sum_r G_r^2(Q^2)\approx {A\over {p+1}} ({{\bar s + M_0^2 - M^2}\over
{Q^2}})^{p+1} 
\label{eleven}
\end {equation}
The first term in the sum on the left-hand-side is just the  elastic nucleon contribution
proportional to its elastic form factor \cite{FF}. Eq.~(\ref{eleven}) is the
integral version of the local inclusive-exclusive relationship \cite{gbw}, a
particular form of which says that the large $Q^2$ behaviour of the nucleon
elastic form factor is given by $(Q^2)^{-(p+1)/2}$ which is in reasonable agreement
with data if $p=3$. From Eq.~(\ref{eleven}), however, one can conclude only that
at least one of the transition form factors must fall like  $(Q^2)^{-(p+1)/2}$;
equivalently, it states that the nucleon elastic form factor cannot fall slower
than this power. This technique does not allow a local version of the inclusive-exclusive
relationship to be derived.

iv) Since, on the right-hand-side of Eq.~(\ref{five}), $T_2$ is to be evaluated in the scaling
region, an alternative approach might be to use the estimate from aymptotic freedom directly.
Unfortunately the canonical light-cone, operator-product machinery \cite{cheng} only gives an
asymptotic estimate for $T_2$ in the unphysical region where $x>1$. Since $\nu T_2$ is a purely
real analytic function there it must be expandable in powers of $x^{-1}$, or in powers of
${\omega '}$:
\begin{equation}
\nu T_2(q^2,\omega ')\approx \sum^{\infty}_{n = 0} c_n(q^2){\omega '}^n \qquad (\omega ' < 1)
\label{thirteen}
\end{equation}
with $c_n(q^2)\approx (\ln -q^2)^{a_n/2b_1}$. Here, $b_1$ is the coefficient of the leading term in the $\beta$-function and $a_n$ the anomolous dimensions of appropiate operators occuring in the expansion of $T_2$. This representation cannot be used in Eq.~(\ref{five}) since that requires $\omega ' > 1$. On the other hand,  $c_n(q^2)$ can be related to the full moments of the structure
functions by expanding the
dispersion relation, Eq.~(\ref{four}), in powers of ${\omega '}$ and comparing coefficients; this is, of course, why the
predictions of asymptotic freedom are expressed in terms of these moments
\cite{georg}. By inverting them, the $q^2$ evolution of $F_2$ can then be determined:
\begin{equation}
F_2(q^2,\omega ') = \int_L {dn\over 2\pi i}{\omega '}^n c(n,q^2)
\label{fourteen}
\end{equation}
Here $c(n,q^2)$ is the analytic continuation of $c_n(q^2)$ to complex values of $m$ and the integration is along a line parallel to the imaginary axis standing to the right of all singularities. Eq.~(\ref{fourteen}) can now be used in the right-hand-side of Eq.~(\ref{eight}) to determine the $q^2$ evolution of the duality relationship. Such a
procedure leads to $A\propto(\ln Q^2)^a$ and $p\approx p_0 + p'\ln\ln Q^2$,
where both $a$ and $p'$ are known, calculable constants in QCD\cite{parisi}. So,
as far as the leading $Q^2$ behaviour is concerned, Eq.~(\ref{eleven}) reads:   
\begin{equation}
\sum_r G_r^2(Q^2)\approx {(\ln Q^2)^a\over ({Q^2})^{p+1 +p'\ln\ln Q^2}} 
\label{twelve} 
\end {equation}
A cautionary note must be added, however: since ${\omega '}^n = e^{n\ln\omega '}$, the $\omega '\rightarrow 1$ behaviour of $F_2$ and, consequently, the  derivation of Eq.~(\ref{twelve}), is sensitive to the large moments ($\vert n\vert\rightarrow\infty$). 
There is, therefore, a delicate question of the uniformity of the expansions which cast possible doubt
on the extrapolation to such small values of $\omega '$ close to $1$. Nevertheless this behaviour is
an intriguing possibility which should be tested phenomenologically. On the other hand,
even if Eq.~(\ref{twelve}) is, in fact, valid it is
unlikely that it can be naively extrapolated to its local form and the QCD
corrections to the elastic form factor thereby inferred \cite {glash}. 

This work was carried out during a productive visit to the INFN at the University of Perugia
in Italy. I would like to thank Claudio Ciofi degli Atti for his kind hospitality and
bringing the renewed interest in the behaviour of the structure functions near $x=1$ to my
attention. I would also like to acknowledge the Department of Energy for its continued
support.


\begin{references}
\bibitem{note} Throughout this paper we use the standard notation in which the
scaling variable $x\equiv Q^2/{2M\nu}$; $q$ is the four-momentum transferred to
the target, which will normally be taken to be a nucleon with mass $M$. For
convenience we define $Q^2 = -q^2$. In the target rest frame the energy transferred
is $p\cdot q/M = q_0\equiv\nu$ where $p$ is the target four-momentum. We also use
$s\equiv (p+q)^2 = M^2 + 2\nu - Q^2$.
\bibitem{finite} If neglected terms of $O(M_0^2/Q^2)$ are kept, they simply
change the lower limit on the integral on the right-hand-side from $1$ to
$(1+M_0^2/Q^2)$.
\bibitem{FF} For the nucleon $G_r(q^2)$ is related to the usual Dirac and
Pauli elastic form factors, $F_{1,2}$, by $G_r^2(q^2)=F_1^2(q^2)+(\mu^2
q^2/4M^2)F_2^2(q^2)$, where $\mu$ is the nucleon magnetic moment.
\end{references}
\end{document}